\documentclass[aip,apl,reprint,showpacs,superscriptaddress]{revtex4-1}
\usepackage[T1]{fontenc}
\usepackage{times}
\usepackage{graphicx,color}
\usepackage{amsfonts,amsmath,amssymb,amsbsy}
\usepackage[colorlinks=true, linkcolor=red, citecolor=red, urlcolor=magenta]{hyperref}

\let\mathbf=\boldsymbol

\def\emph#1{\textcolor{blue}{#1}}
\begin{document}

\title{Manipulation of magnetic skyrmions in a locally modified synthetic antiferromagnetic racetrack}

\author{R. P. Loreto}
\thanks{These authors contributed equally to this work.}
\affiliation{Departamento de F\'{i}sica, Universidade Federal de Vi\c{c}osa, Vi\c{c}osa, Minas Gerais 36570-900, Brazil}

\author{X. Zhang}
\thanks{These authors contributed equally to this work.}
\affiliation{School of Science and Engineering, The Chinese University of Hong Kong, Shenzhen, Guangdong 518172, China}

\author{Y. Zhou}
\affiliation{School of Science and Engineering, The Chinese University of Hong Kong, Shenzhen, Guangdong 518172, China}

\author{M. Ezawa}
\affiliation{Department of Applied Physics, The University of Tokyo, 7-3-1 Hongo, Tokyo 113-8656, Japan}

\author{X. Liu}
\affiliation{Department of Electrical and Computer Engineering, Shinshu University, 4-17-1 Wakasato, Nagano 380-8553, Japan}

\author{C. I. L. de Araujo}
\email[E-mail:~]{dearaujo@ufv.br}
\affiliation{Departamento de F\'{i}sica, Universidade Federal de Vi\c{c}osa, Vi\c{c}osa, Minas Gerais 36570-900, Brazil}

\begin{abstract}
In skyrmion-based racetrack memories, the information encoded by skyrmions may be destroyed due to the skyrmion Hall effect, which can be surmounted by using synthetic antiferromagnetic racetracks. Hence, the manipulation of skyrmions in synthetic antiferromagnetic racetracks is important for practical applications. Here, we computationally study the interaction between a pair of skyrmions and a locally modified region in a synthetic antiferromagnetic racetrack, where the perpendicular magnetic anisotropy or thickness is locally adjusted to be different from that of the rest region of the racetrack. It is found that the skyrmions can be attracted, repelled, and even trapped by the locally modified region in a controllable manner. Besides, we demonstrate that the skyrmion location can be precisely determined by the locally modified region. The possible manipulation of skyrmions by utilizing locally modified regions in a synthetic antiferromagnetic racetrack may be useful for future skyrmion-based applications.
\end{abstract}

\date{7 March 2019}

\maketitle


Racetrack memories~\cite{Parkin_SCIENCE2008,Parkin_NNANO2015} have been extensively investigated due to their ultra-high information processing speed and low power consumptions, in comparison with memory technologies based on magnetization switching induced by Oersted fields and Joule effects~\cite{Araujo_PRApplied2016}.
In a conventional racetrack memory, the information is written by creating magnetic domain walls through local magnetization switching, which can be realized by harnessing the effect of spin-transfer torque (STT)~\cite{STT2008}.
However, the information carried by domain walls may be destroyed or lost due to the domain wall susceptibility to imperfections at racetrack borders.
Therefore, the racetrack memory with magnetic skyrmions acting as information carriers has been proposed to provide a potential route to overcome information loss caused by non-desired impurities and defects~\cite{Fert_NNANO2013,Sampaio_NNANO2013,Iwasaki_NNANO2013,Tomasello_SREP2014,Xichao_SREP2015A}.

Magnetic skyrmions are nanoscale topological spin textures~\cite{Nagaosa_NNANO2013,Wiesendanger_Review2016,Fert_NATREVMAT2017,Wanjun_PHYSREP2017}, which usually can be found in magnetic materials with asymmetric or frustrated exchange interactions~\cite{Roszler_NATURE2006,Muhlbauer_SCIENCE2009,Leonov_NCOMMS2015,Zhang_NCOMMS2017}. A number of theoretical and experimental works have shown that magnetic skyrmions can be used as building blocks for racetrack memories~\cite{Kang_PIEEE2016,Finocchio_REVIEW2016,Bhatti_MAT2017,Koshibae_JJAP2015,Du_NCOMMS2015,Woo_NCOMMS2018}, logic computing devices~\cite{Xichao_SREP2015B}, and other applications~\cite{Li_NANO2017,Yang_NANO2017,Bourianoff_AIPADV2018,Prychynenko_PRApplied2018,Nozaki_APL2019}. However, it is difficult to create single isolated skyrmions on the racetrack and avoid the skyrmion trajectory deviation due to the skyrmion Hall effect~\cite{Zang_PRL2001,Wanjun_NPHYS2017,Litzius_NPHYS2017}. Several works have been performed in order to overcome these obstacles for practical skyrmion-based racetrack-type applications.
For example, the generation of skyrmions on the track have been achieved with utilization of different approaches, such as transforming skyrmions from domain walls~\cite{Yan_NCOMMS2014,Wanjun_SCIENCE2015,Yu_NL2017}, creating skyrmions from notches~\cite{Iwasaki_NNANO2013,Buttner_NNANO2017}, and creating skyrmions by unique electric pulses~\cite{Woo_NE2018}.

On the other hand, in order to avoid the detrimental effect of the skyrmion Hall effect, which prevents skyrmions from moving along the central line of the racetrack, several methods have been proposed recently. For examples, the racetrack modification by insertion of lateral stripes with higher perpendicular magnetic anisotropy (PMA)~\cite{Purnama_SREP2015,Lai_SREP2017}, the utilization of ferromagnetic skyrmioniums~\cite{Xichao_PRB2016A}, the utilization of antiferromagnetic (AFM) skyrmions~\cite{Xichao_SREP2016,Tretiakov_PRL2016} and synthetic antiferromagnetic (SyAF) skyrmions~\cite{Xichao_NCOMMS2016,Xichao_PRB2016B}. Since the topological charge is zero for AFM and SyAF skyrmions, they can strictly move along the central line of the racetrack, showing no skyrmion Hall effect~\cite{Xichao_SREP2016,Tretiakov_PRL2016,Xichao_NCOMMS2016,Xichao_PRB2016B}.

In this work, we numerically investigate the controllable manipulation of skyrmions in a SyAF racetrack by locally modifying the PMA under the framework of micromagnetics.
Such an investigation is necessary for further applications and can be used for controlling the spacing between adjacent skyrmions, defining the bit length, where a magnetic tunnel junction (MTJ) could be placed to detect the skyrmion, defining the bits ``0'' or ``1'' whether the skyrmion is present or absent, and protecting skyrmions from external fluctuations such as stray fields from nearby racetracks, similar to what have been proposed and studied for domain wall-based racetracks~\cite{Parkin_SCIENCE2008}.
We use a protocol to create antiferromagnetically coupled N{\'e}el-type skyrmions with perpendicular spin-polarized current applied on the racetrack, as recently demonstrated by Zhang et al.~\cite{Xichao_NCOMMS2016,Xichao_PRB2016B}. We then investigate the longitudinal motion of SyAF skyrmions driven by pulses of spin-polarized current on the racetrack with a locally modified region.
Our simulation results suggest that it is effective to control and manipulate skyrmions in SyAF racetracks by constructing locally modified regions with different thickness or PMA strength. 

\begin{figure}[t]
\centerline{\includegraphics[width=0.48\textwidth]{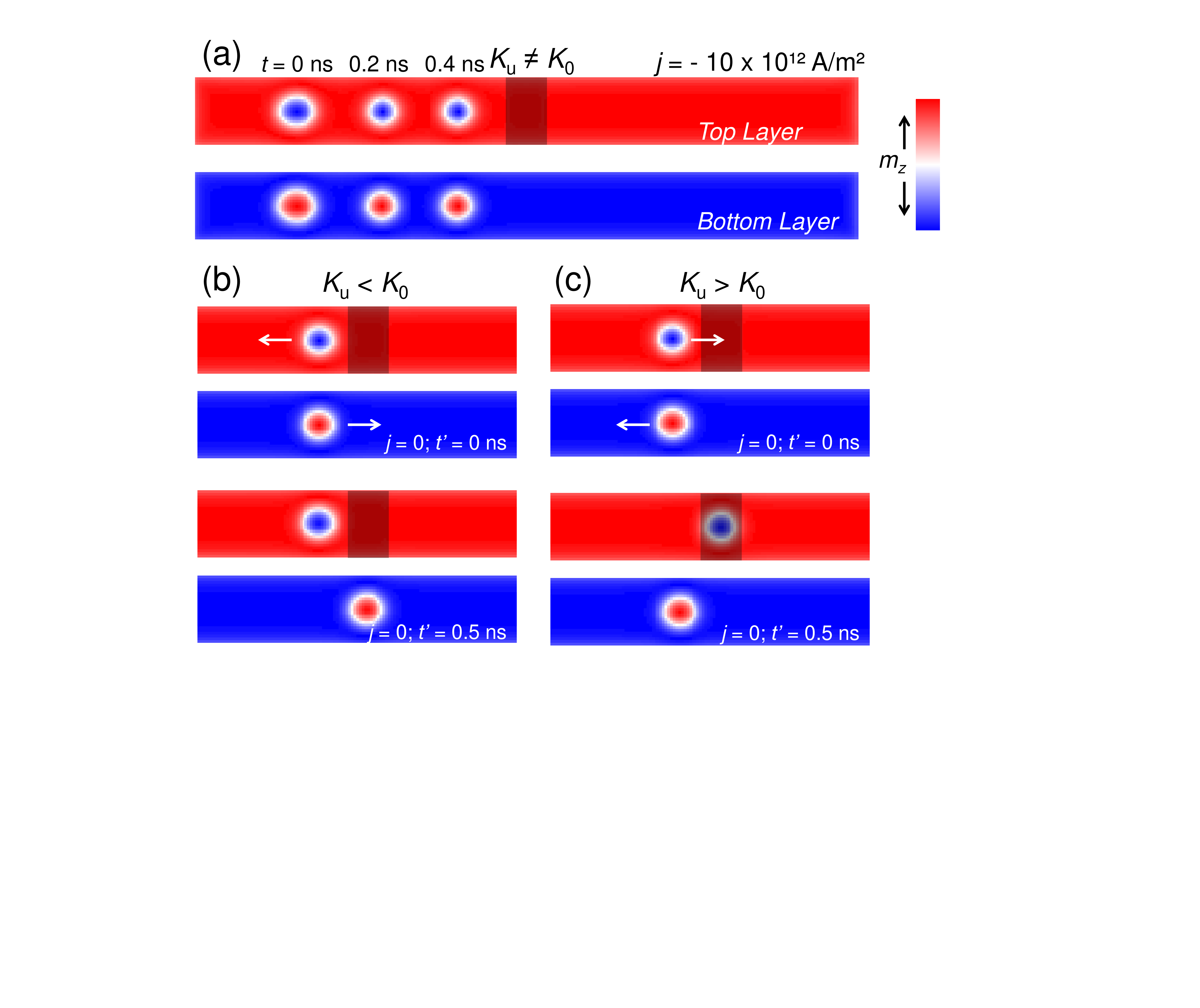}}
\caption{%
(\textbf{a}) A SyAF racetrack with $30$-nm-long central region modified to have PMA constant $K_{\text{u}}$ different from that of the rest region $K_{\text{0}}$. The SyAF skyrmions move around $200$ nm in $0.4$ ns before the applied current of $j = 10 \times 10^{12}$ A m$^{-2}$ is switched off.
(\textbf{b}) The top-layer skyrmion in the locally modified racetrack is expelled when the central region has $K_{\text{u}}<K_{\text{0}}$, while the bottom-layer skyrmion is pinned and decoupled from the top-layer skyrmion.
(\textbf{c}) The top-layer skyrmion is pinned in the locally modified region when $K_{\text{u}}>K_{\text{0}}$, while the bottom-layer skyrmion is repelled and decoupled from the top-layer skyrmion.
See Supplementary Movie.}
\label{FIG1}
\end{figure}


\begin{figure}[t]
\centerline{\includegraphics[width=0.48\textwidth]{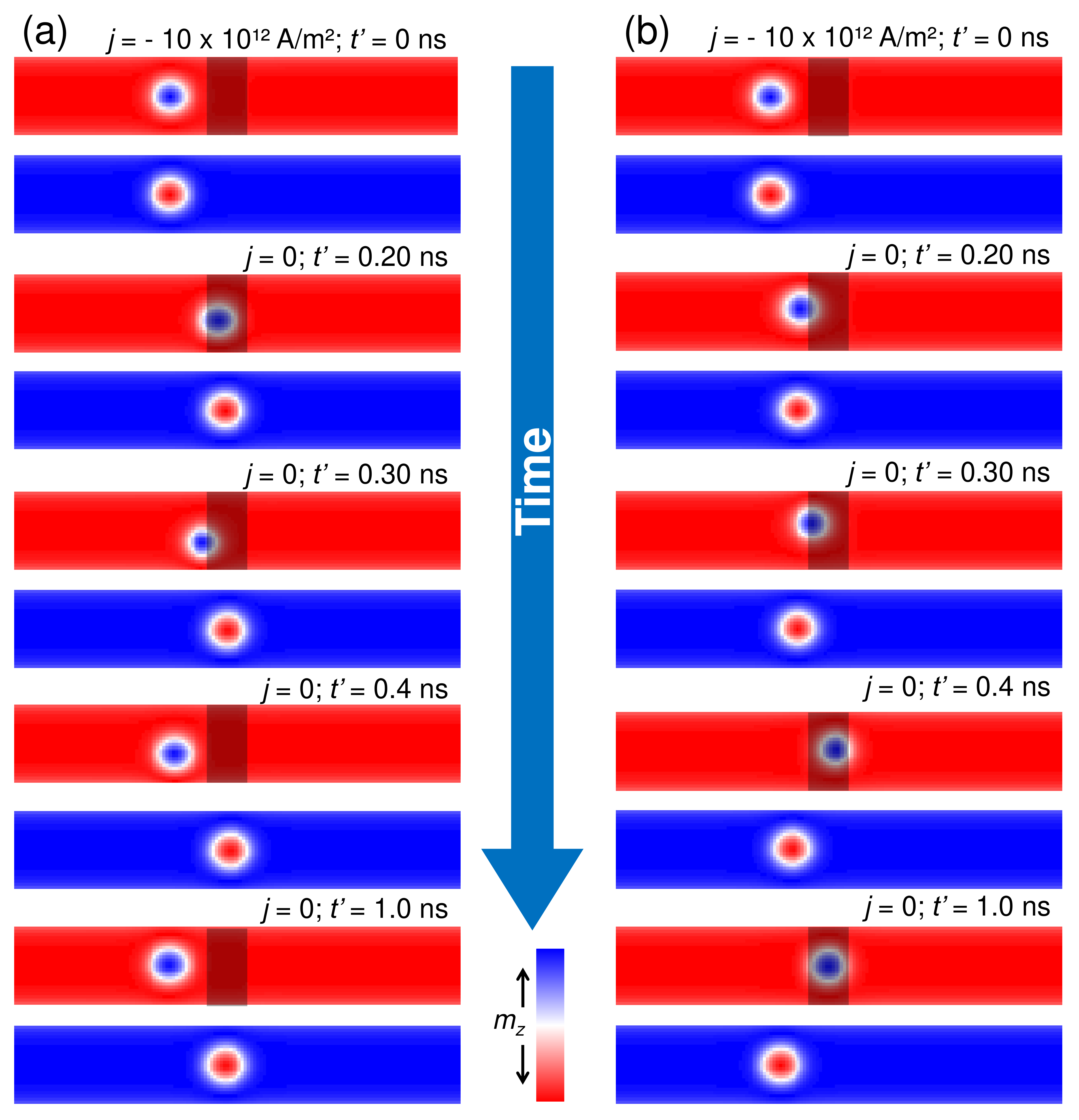}}
\caption{%
(\textbf{a}) A SyAF skyrmion is presented near the thicker central region (indicated by shaded boxes) just after the application of a spin-polarized current. The top-layer and bottom-layer skyrmions are decoupled and the top-layer skyrmion is repelled for a distance about $30$ nm at $j=0$, while the bottom-layer skyrmion is pinned just under the locally modified region.
(\textbf{b}) A SyAF skyrmion stops near the thinner central region (indicated by shaded boxes) of the top racetrack, it is decoupled with the top-layer skyrmion pinned in the center of the locally modified region while the bottom-layer skyrmion is repelled around $30$ nm at $j=0$.
See Supplementary Movie.
}
\label{FIG2}
\end{figure}

\begin{figure}[t]
\centerline{\includegraphics[width=0.48\textwidth]{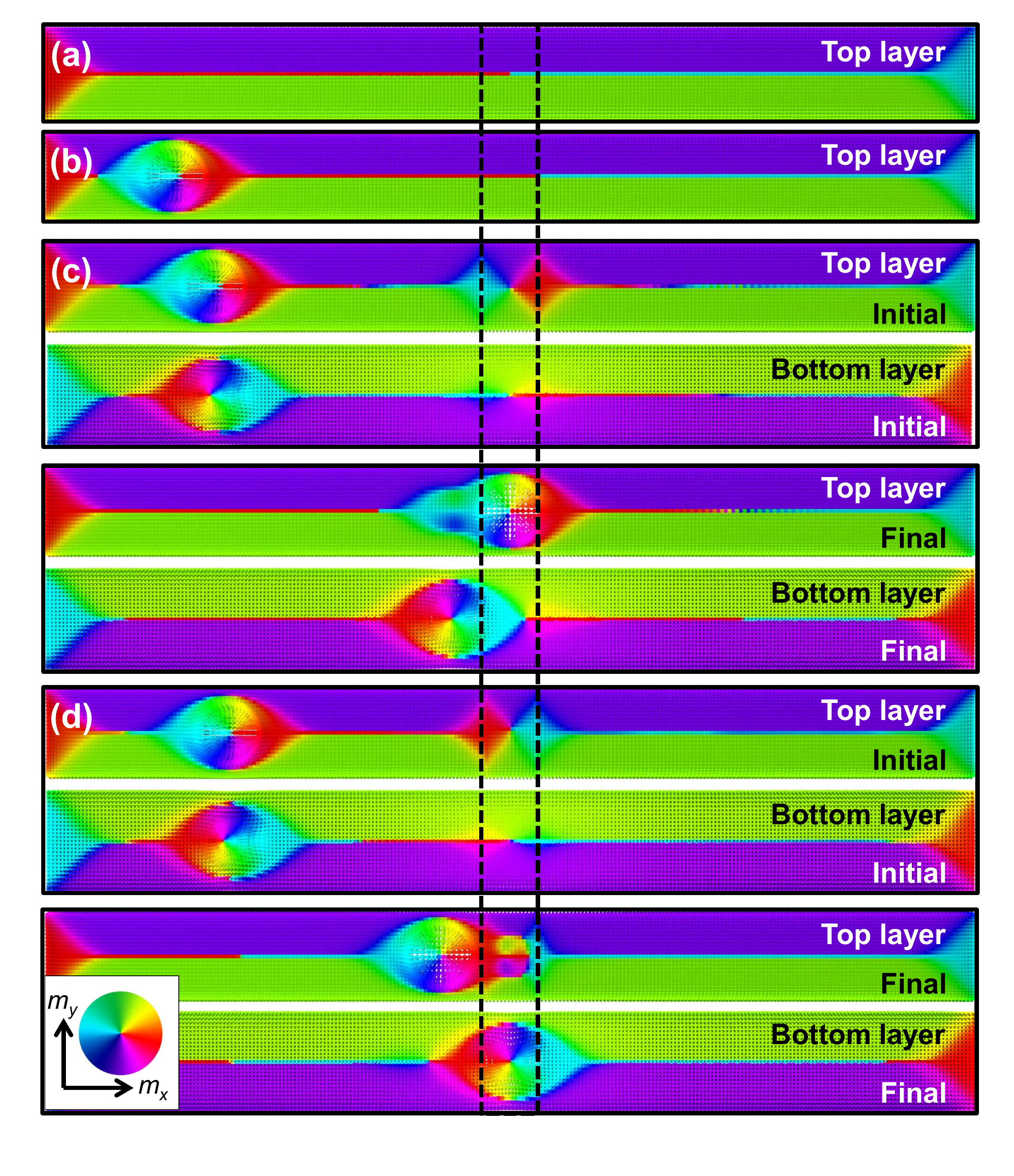}}
\caption{%
Top views of the in-plane magnetization configurations of the SyAF racetrack.
(\textbf{a}) The ground state of the unmodified SyAF racetrack ground state where in-plane magnetization configurations are naturally formed. Only the top layer is shown.
(\textbf{b}) The unmodified racetrack including a SyAF skyrmion. Only the top layer is shown.
(\textbf{c}) The locally modified racetrack including a SyAF skyrmion, where the central region between the dashed lines has a thinner thickness and thus behaving as higher PMA.
(\textbf{d}) The locally modified racetrack including a SyAF skyrmion, where the central region between the dashed lines has a thicker thickness and thus behaving as lower PMA.
The formation of extra in-plane magnetization configurations in the locally modified region leads to the skyrmion pinning and expulsion due to the dipole-dipole interaction. The skyrmion-like structure that appears inside the modified region on (\textbf{d}) is due to the skyrmion in the bottom layer, like an imprinted texture.}
\label{FIG3}
\end{figure}

\begin{figure}[t]
\centerline{\includegraphics[width=0.48\textwidth]{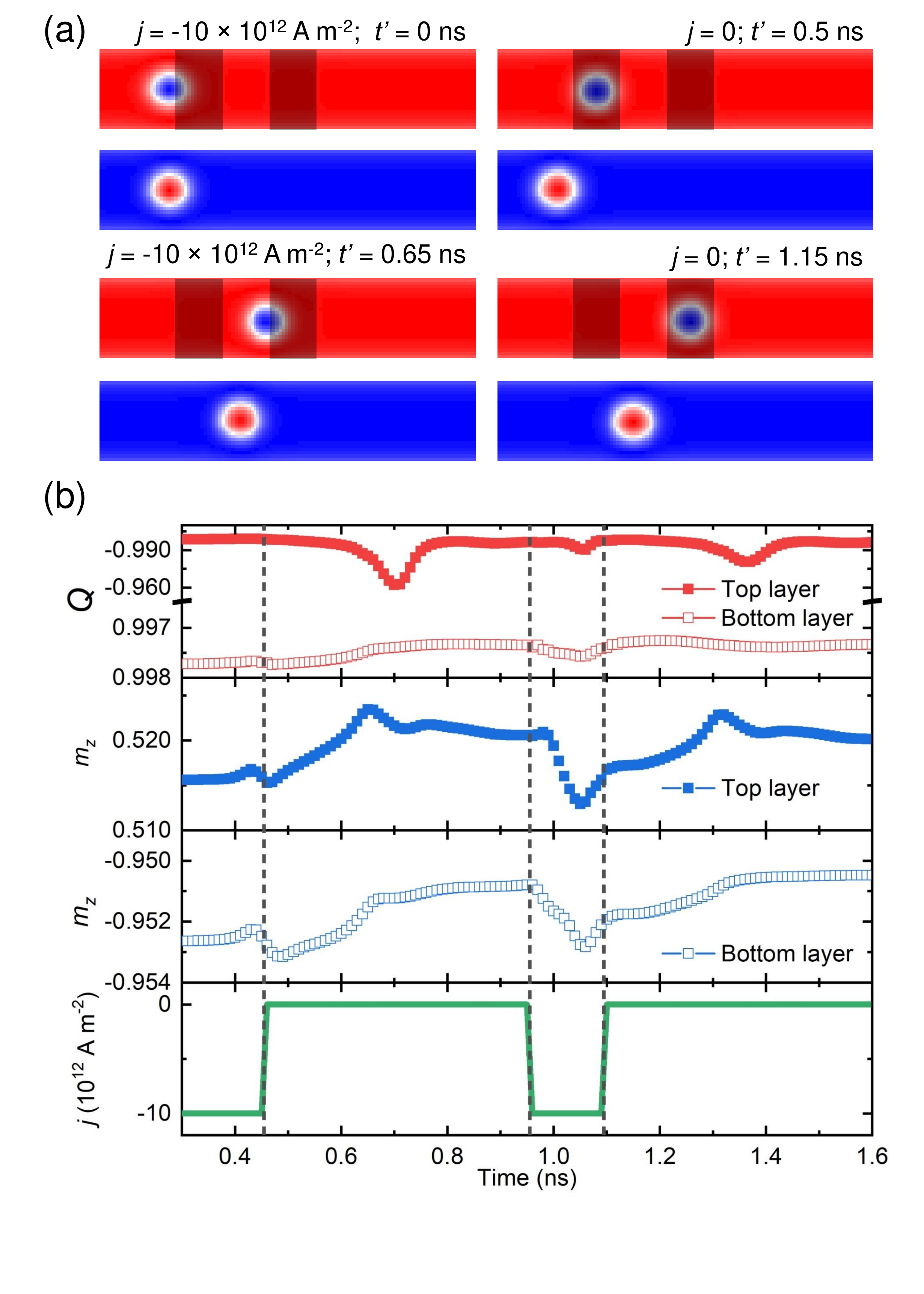}}
\caption{%
(\textbf{a}) Evolution of the skyrmion pinning in the top layer of a SyAF racetrack, which demonstrates the utilization of locally modifications on the racetrack (indicated by shaded boxes) to form successive bits with length of $30$ nm separated by $30$ nm. The top-layer skyrmion is accommodated in each bit in $0.5$ ns after the application of spin-polarized current pulses.
(\textbf{b}) Time evolution of the topological charge $Q$, reduced out-of-plane magnetization $m_z$, and applied current density $j$ corresponding to (\textbf{a}). The dashed lines are a guide to the eye for the instants when the current changes value. Please note that, in the continuous theory, the topological charge for a ground-state skyrmion is $Q = \pm 1$.
The magnetization near the edges of the modified region are tilted due to the DMI. Since these tilted magnetization also make a contribution to the topological charge, the summation of topological charges in bottom and top layers is not exactly zero in the numerical calculation.
}
\label{FIG4}
\end{figure}


As shown in Fig.~\ref{FIG1}(a), a spin-polarized current of $j=10 \times 10^{12}$ A m$^{-2}$ is first applied to drive the SyAF skyrmions into motion, moving about $200$ nm in $0.4$ ns and stop near a $30$-nm-long central region of the SyAF racetrack top layer (highlighted in shaded boxes). The $30$-nm-long central region is modified to have a higher or lower PMA constant $K_{\text{u}}$ than that of the rest region of the racetrack $K_{\text{0}}$.
The default PMA constant of the racetrack $K_{\text{0}}=0.8\times10^6$ J m$^{-3}$ is defined based on the experimental CoPt layers~\cite{Sampaio_NNANO2013} and used for both top and bottom layers, while PMA constant of the locally modified region $K_{\text{u}}=0.6\times10^6$ and $2.0\times10^6$ J m$^{-3}$  and it can be realized by using different materials~\cite{Purnama_SREP2015,Lai_SREP2017}.

Figure~\ref{FIG1}(b) shows the dynamic behavior of a SyAF skyrmion close to the locally modified region with $K_{\text{u}}<K_{\text{0}}$ under zero current (i.e., $j=0$ A m$^{-2}$).
The asymmetry of the PMA in the top layer of the racetrack leads to the motion as well as decoupling of the SyAF skyrmion, even at $j=0$ A m$^{-2}$. The decoupling of the SyAF skyrmion depends on the strength of the interlayer coupling and may not occur if the relative AFM interlayer coupling is higher than 10$^{-4}$ (see Supplementary Material).
The top-layer skyrmion is repelled from the locally modified region about $20$ nm in $1.5$ ns, while the bottom-layer skyrmion is pinned in the locally modified region.
A similar phenomenon is observed when the locally modified region has a higher PMA constant, namely, $K_{\text{u}}>K_{\text{0}}$, as shown in Fig.~\ref{FIG1}(c), where the top-layer skyrmion is pinned in the locally modified region and the bottom-layer skyrmion is repelled from the locally modified region after the decoupling of the SyAF skyrmion.
It should be mentioned that there is no skyrmion Hall effect when the skyrmions are decoupled since $\alpha = \beta$ is adopted for simplicity, as discussed in Ref. \onlinecite{Xichao_IEEE2017}. In experiments, it could be difficult to realize the material condition of $\alpha = \beta$, which means the skyrmion Hall effect may lead to the decoupling of top-layer and bottom-layer skyrmions. Therefore, a stronger AFM interlayer coupling is usually preferred for experimental samples. On the other hand, if the driving force provided by the spin-polarized current is larger than a certain threshold when $\alpha \neq \beta$, the top-layer and bottom-layer will also be decoupled. Hence, it is important to control the magnitude of the injected current density in experiments.

The results given in Fig.~\ref{FIG1} indicates the possibility of manipulating skyrmions by locally modifying the PMA constant in a racetrack, which could be practically realized by using different materials in a well defined region. 
However, the experimental realization of such a locally modified region demands several steps for the racetrack fabrication. The other possible method is to induce a shape anisotropy that emulates the difference in the PMA constant.
In this way, we investigate the dynamic behavior of a SyAF skyrmion near the locally modified region, where the thickness of this region is tuned to be thicker or thinner than the rest region of the racetrack.
Such a method has more compatibility with experimental nano-fabrication process, because it demands less  lithographic steps and since an ion milling step can be applied to create regions on the racetrack developed with the same material but with different thickness.

Figure~\ref{FIG2}(a) shows the response of a SyAF skyrmion to a locally modified region, where the thickness is $0.5$ nm thicker than the rest region of the racetrack. It is found that the top-layer skyrmion is expelled about $20$ nm in $0.5$ ns far from the central locally modified region, while the bottom-layer skyrmion is pinned just below the thicker region. Here we find that the thicker region is able to mimic the effect of a locally modified region with lower PMA constant, in this case the antiferromagnetic coupling is weaker, making the anisotropy smaller and it corresponds to the case given in Fig.~\ref{FIG1}(b), but with a higher efficiency, as the top-layer skyrmion moves about the same distance $3$ times faster. Successively, the decrease in the racetrack thickness is studied by locally modifying the central region to be $0.5$ nm thinner than the rest region of the racetrack, opposite to the previous case, the AFM coupling is stronger in the thinner region, behaving as higher value of anisotropy.
As shown in Fig.~\ref{FIG2}(b), when a SyAF skyrmion is close to the thinner central locally modified region ($t = 0$ ns), the top-layer skyrmion is attracted to the central of the modified region ($t = 0.5$ ns), while the bottom-layer skyrmion is a bit expelled from the locally modified region.
In order to understand why the local modifications of PMA constant and shape anisotropy (i.e., thickness) in the racetrack are able to induce the SyAF skyrmion decoupling and pin or expel top-layer/bottom-layer skyrmions, we continue to investigate the in-plane magnetization configurations of the SyAF racetrack.

Figure~\ref{FIG3} shows the in-plane magnetization configurations for different steps, which are very small and not noticeable in Fig.~\ref{FIG1} and Fig.~\ref{FIG2} presenting the three-dimensional magnetization.
The in-plane magnetization configuration of the ground state of the racetrack is given in Fig.~\ref{FIG3}(a), and the in-plane magnetization configuration of the relaxed racetrack including a N{\'e}el-type SyAF skyrmion is given in Fig.~\ref{FIG3}(b).
As shown in Fig.~\ref{FIG3}(c), for the locally modified region with $K_{\text{u}}>K_{\text{0}}$ or a thinner thickness, two in-plane magnetization areas form in the borders of the locally modified region of the top racetrack, which match the in-plane magnetization configuration of the top-layer skyrmion. Similar behavior is observed in the bottom racetrack, however, the formed in-plane magnetization areas don't match the in-plane magnetization configuration of the bottom-layer skyrmion. Hence, when the top-layer skyrmion is close to the border of the locally modified region, it is attracted by dipole-dipole interaction (DDI) and pinned by Heisenberg exchange interaction, while the bottom-layer skyrmion is expelled by the same interactions.
Figure~\ref{FIG3}(d) shows the in-plane magnetization configuration for the locally modified region with $K_{\text{u}}<K_{\text{0}}$ or a thicker thickness. Here, the in-plane magnetization configurations at the borders of the locally modified region are opposite to that observed in Fig.~\ref{FIG3}(c), therefore, following the same principle the top-layer skyrmion is expelled from the locally modified region and the bottom-layer skyrmion is attracted and pinning by the locally modified region.

The presented results suggest that it is possible to apply local modifications in the racetrack to generate bits, where the skyrmion is pinned, and in that position a MTJ could be fabricated to sense the skyrmion for information read-out processing.
In order to test the shortest bit length we apply two separated spaces where the thickness of the racetrack is reduced by $0.5$ nm.
As shown in Fig.~\ref{FIG4}(a), we demonstrate that the skyrmion can travel between two bits in a controllable manner, where the locally modified regions are able to keep the skyrmion in a desired location for read-out measurement using MTJ~\cite{Loreto_JMMM2017}.
Figure~\ref{FIG4}(b) shows very small fluctuations in both top-layer and bottom-layer skyrmion topological charges during their motion between the bits, indicating the locally modified region has tiny influence on the skyrmion structure.
It is worth mentioning that a single SyAF racetrack can have multiple bits (i.e. skyrmions), however, it is important to control the spacing between neighboring skyrmions in order to avoid skyrmion-skyrmion interaction~\cite{Xichao_SREP2015A}.


In conclusion, we have investigated the dynamic behavior of skyrmions near the locally modified region of a SyAF racetrack, where the PMA constant or shape anisotropy is artificially adjusted.
We have shown that it is possible to use the locally modified region to manipulate the skyrmion, such as defining the bit length and setting the right location to develop the read-out MTJ.
We found that certain in-plane magnetization configurations are generated at the borders of the locally modified region, which can interact with the skyrmion.
The SyAF skyrmion could be decoupled into a top-layer skyrmion and a bottom-layer skyrmion when it is close to the locally modified region, due to the DDI and exchange interaction between skyrmions and in-plane magnetization configurations at the borders of the locally modified region. Depending on the PMA constant or thickness of the locally modified region in relation to the rest region of the racetrack, the skyrmions can be either attracted or expelled by the locally modified region.
We have also demonstrated that by using locally modified regions with thinner thickness and with small separation among each other, it is possible to move skyrmions between desired bit locations with very small driving current pulse.

\section*{Methods} 

The micromagnetic simulations are performed with the GPU-accelerated micromagnetic simulator $\textsc{MuMax3}$, which solves the Landau-Lifshitz-Gilbert (LLG) equation augmented with the adiabatic STT~\cite{MUMAX},

\begin{align}
\label{eq:LLG}
\frac{\partial \boldsymbol{m}}{\partial t}=&-\gamma\boldsymbol{m}\times\boldsymbol{H}_\text{eff}+\alpha\left(\boldsymbol{m}\times\frac{\partial\boldsymbol{m}}{\partial t}\right) \\ \notag
&+\frac{\gamma\hbar jP}{2e\mu_0 M_{\text{S}}}\left(\boldsymbol{m}\times\frac{\partial\boldsymbol{m}}{\partial x}\times\boldsymbol{m}\right) \\ \notag
&-\frac{\beta\gamma\hbar jP}{2e\mu_0 M_{\text{S}}}\left(\boldsymbol{m}\times\frac{\partial\boldsymbol{m}}{\partial x}\right),
\end{align}
where $\boldsymbol{m} = \boldsymbol{M}/M_\text{S}$ is the reduced magnetization, $M_\text{S}$ is the saturation magnetization, $\gamma$ is the gyromagnetic ratio with absolute value, $\alpha$ is the Gilbert damping parameter, $\beta$ is the non-adiabatic STT strength, $\hbar$ is the reduced Planck constant, $\mu_0$ is the vacuum permeability, $e$ is the electron charge, $P$ is the polarization ratio of the electric current, and $\boldsymbol{H}_\text{eff}$ is the effective field. The considered micromagnetic energy terms include the interlayer and intralayer Heisenberg exchange interaction energies, Dzyaloshinskii-Moriya interaction (DMI) energy, PMA energy, applied magnetic field energy, and DDI energy. The energy density of the DMI can be expressed as
\begin{align}
\label{eq:energy}
E_\text{DMI} = \frac{D}{M^{2}_{s}} (M_z \partial_x M_x + M_z \partial_y M_y \\ \notag
- M_x \partial_x M_z - M_y \partial_y M_z ).
\end{align}
The third and fourth terms on the right-hand side of Eq.~\eqref{eq:LLG} is related to the adiabatic and non-adiabatic STTs provided by the spin-polarized current density $\boldsymbol{j}$ in the racetrack.

The simulated racetrack is composed by two ferromagnetic layers with dimensions of $500\times50\times2$ nm$^3$, which are separated by a spacer of $1$-nm-thick Ru layer to mimic a SyAF multilayer.
The magnetic parameters used in our micromagnetic simulations are adopted from Refs.~\onlinecite{Sampaio_NNANO2013,Yan_NCOMMS2014,Xichao_NCOMMS2016,Xichao_PRB2016B,Xichao_PRB2016A} as follows:
the saturation magnetization $M_{\text{S}} = 5.8 \times 10^5$ A m$^{-1}$; exchange stiffness $A_{\text{ex}} = 15 \times 10^{-12}$ J m$^{-1}$; relative AFM interlayer exchange coupling varying from $-1$ to $-10^{-12}$ J m$^{-1}$; the DMI constant equals $D = 3.5 \times 10^{-3}$ J m$^{-2}$; the PMA constant $K_{0} = 0.8 \times 10^{6}$ J m$^{-3}$, the damping parameter $\alpha = 0.3$; the non-adiabatic STT parameter $\beta = 0.3$, and the spin polarization factor $P = 0.4$.

The images of magnetization configurations are obtained by using the $\textsc{MuView}$ software and the topological charges are calculated based on the following definition~\cite{Wanjun_PHYSREP2017}
\begin{equation}
Q=\frac{1}{4\pi}\int\boldsymbol{m}\cdot(\frac{\partial\boldsymbol{m}}{\partial x}\times\frac{\partial\boldsymbol{m}}{\partial y})dxdy.
\label{eq:Q}
\end{equation}
In theory, the topological charge $Q$ for a skyrmion is strictly equal to $\pm 1$ in the continuous model, however, due to the numerical discretization of the micromagnetic simulation process, we have $Q\approx \pm 1$ for a skyrmion in the micromagnetic system.

\section*{Acknowledgments}
This work was partially supported by CAPES (Finance Code 001), CNPq and FAPEMIG (Brazilian agencies).
X.Z. acknowledges the support by the Presidential Postdoctoral Fellowship of The Chinese University of Hong Kong, Shenzhen (CUHKSZ).
Y.Z. acknowledges the support by the President's Fund of CUHKSZ, the National Natural Science Foundation of China (Grant No. 11574137), and Shenzhen Fundamental Research Fund (Grant Nos. JCYJ20160331164412545 and JCYJ20170410171958839).
M.E. acknowledges the support by the Grants-in-Aid for Scientific Research from JSPS KAKENHI (Grant Nos. JP18H03676, JP17K05490, JP15H05854), and also the support by CREST, JST (Grant No. JPMJCR16F1).



\vbox{}
\noindent\textbf{Author Contributions}

\noindent
C.I.L.A coordinated the project.
R.P.L. and C.I.L.A performed the simulation.
C.I.L.A, R.P.L. and X.Z. prepared the manuscript with input from Y.Z., M.E. and X.L.
All authors discussed the results and reviewed the manuscript.

\vbox{}
\noindent\textbf{Competing Financial Interests}

\noindent
The authors declare no competing financial interests.

\end{document}